\documentstyle[11pt,newpasp,twoside,epsf]{article}
\def\edcomment#1{\iffalse\marginpar{\raggedright\sl#1\/}\else\relax\fi}
\reversemarginpar

\begin{document}
\title{Modern Stellar Evolution Models for Low Mass Stars}

\author{Vittorio Castellani}
\affil{INAF, Osservatorio Astronomico di Roma, Via di Frascati 33,
00040 Monteporzio}

\begin{abstract}
The status of the art for evolutionary models of low mass,
population II stars is revisited, stressing the need for the
models to be preliminarily tested with suitable observational
data. The uncertainties still affecting the theory are discussed
in the aim of guiding the users in a reasoned choice of the
models. The powerful coupling of evolutionary with pulsational
theories is shortly recalled. A short conclusion and a mention of
the problem of tilted HB close the paper.

\end{abstract}

\section{Introduction}

When I was invited to give a talk about "Modern Stellar Evolution
Models"  I began wondering about the meaning of the word "modern".
Obviously, with such a word the promoters of the meeting intended
refer to an increased reliability reached by recent stellar
evolutionary models: thus "modern" would mean "what we are  able
to do at the present time".  However, stellar models appearing in
the current literature not necessarily and not always represent
the best ones can we now do, i.e., not necessarily "recent" means
"modern". Bearing in mind such a warning, in this talk I will
discuss the "status of the art" of stellar evolutionary theories,
highlighting which -at least in my opinion - should be the more
suitable ground  for modern stellar models.

In order to approach this issue, one has to remind that stellar
evolution theories were born to account for the evolutionary
evidences as given by the CM diagram of stellar clusters. In
particular, evolutionary theories for low mass stars  find their
experimental (observational) counterpart in the CM diagrams of
Galactic Globular Clusters (GGC). It may be interesting,
especially for the younger people in this audience,  to recall how
tremendous was the increase of observational evidences concerning
GGCs along the past century. Before the II World War only Red
Giant Branches (RGB) and  Horizontal Branches (HB) were barely
known, whereas the first evidence for the bright end of the
cluster Main Sequence was only attained during the 50's (Arp, Baum
\& Sandage 1953).

Since that time, and for several decades, a huge amount of
investigations  dealt with such a "canonical portion" of the
cluster CM diagram, lacking any observational evidence for the
fainter stars we knew should populate the cluster. Only in rather
recent time the improved capability of both ground and space based
observations allowed us to gain evidence on the large majority of
cluster stars, adding information on the "newcomers" as
represented by the  Very Low Mass (VLM) MS stars, as well as
reaching the bright end of the  cooling White Dwarf sequence,
fainter WDs waiting for deeper observations.

In the same century, following the pioneering work of the "Fathers
of Evolution" (Hayashi, Schwarzschild, followed by Kippenhan,
Iben, Demarque ....) stellar models were continuously improved in
order to follow the evolution of low mass stars all along their
nuclear burning phases and down to their final cooling as WD. As a
result, about twenty years ago stellar models were already able to
"reasonably" account for the main CM diagram features, where
"reasonably" means within observational uncertainties concerning
not only the photometry but also the cluster reddening and
distance modulus. In this way, the main goal of evolutionary
theories was achieved, i.e., to account for the observed CM
diagram morphologies.

\section{Stellar models and theoretical uncertainties}

In normal life, theories are expected to fit experimental data.
However, the quoted agreement between evolutionary theories and
observations has early prompted several people to use theoretical
results in a rather unconventional way, attempting a theoretical
calibration of observational data. As an example, by comparing the
observed with the predicted luminosity of the HB in a given
globular, one could derive the distance modulus of the cluster and
- coming back to theory - its age. Similar procedures have however
raised several debates for the very simple reason that  at that
time different authors gave (not surprisingly, as we will see)
different predictions for the HB luminosity as well as for other
relevant observational parameters.

To  have light on such an occurrence, one has to make clear a
relevant point, not generally acknowledged by  people not
concerned with stellar models. "Model" is indeed an ambiguous
word, suggesting perhaps some degree of freedom. On the contrary,
a stellar model is a strong and solid computational architecture,
without any degree of freedom, and a stellar model is as good as
the input physics used to feed the model. Different models with
the same input physics must give, as they give,  the same results.
If not, it is matter of (rare) computational mistakes.

What I am saying is that different results were, and are, the
expected consequence of different input physics. The amount of
physics required by a model is indeed tremendous, with difficult
quantitative evaluations of the behavior of stellar plasma, which
normally requires some sort of approximation and that along the
time has been subject to continuous improvements. To avoid an
unnecessary and damaging confusion, people computing stellar
models should update their code to the most trustworthy and
documented physical evaluations, submitting the result to all
possible observational tests. This, in my feeling, appears the
only reasonable route to modern stellar models. At the same time,
when fitting observations, models should be chosen on the basis of
a proved adequacy of the adopted physics, and not according to
criteria of convenience, institution, friendship .. if not
nationality.

\section{The Solar Connection}

To test the solidity of the various alternative model results, and
of the related physical assumptions,  one needs more stringent
observational constraints than given by globular cluster stars. A
first  opportunity is offered by the SUN, since helioseismology
already provided relevant constraints on its internal structure.
The Sun is indeed a low mass stars, which - like GGC stars - will
undergo electron degeneracy during the H shell burning evolution,
evolving as a RG till attaining its central He burning, HB phase.
Thus, the Sun represents a reasonable test for low mass, central H
burning structures, even if with a different metallicity.

Constraints from the solar structure have already required several
improvements in the previous input physics as given, e.g.,  by the
OPAL results concerning the Equation of State (EOS) for stellar
matter and the evaluation of radiative opacities, bringing also
the additional evidence for the efficiency of  element diffusion
in the external solar layers. Adopting such a modern evolutionary
scenario, one eventually attains quite a good agreement of
theoretical Solar Standard Models (SSM) with solar constraints, as
shown in fig. 1, where theoretical predictions on the behavior of
the ratio between pressure and density along the solar structure
are compared with helioseismic constraints.

\begin{figure}
\plotfiddle{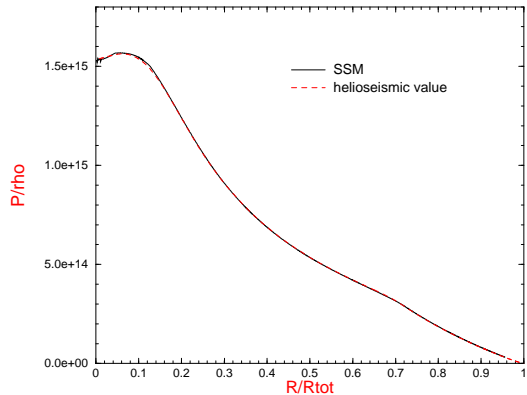}{5.cm}{0}{40}{40}{-110}{-30}
\caption{The predicted run of P/$\rho$ along the solar structure
(full line) as compared with the corresponding helioseismic
experimental data (dashed line).}
\end{figure}

As a relevant points, one finds that the input physics passing the
solar test provides stellar models which appear also in agreement,
even in varying the star metallicity,  with the absolute CM
diagram of stars in nearby open clusters with distance moduli from
the Hipparcos parallaxes (Castellani et al 2001). Waiting for
further observational constraints, as expected from the next
generation of astrometric satellites, in my opinion one should
regard as MODERN stellar models only those models passing both the
solar and the Hipparcos tests.

However, these tests are largely neglected, making difficult an a
priori choice among different results. Bearing this in mind, in
the following we will discuss the "status of the art" from the
recent literature, to make clear the existence of additional
uncertainties and to help people in properly handling theoretical
data.

\section{Stellar Models for Low Mass Stars: the canonical portion}

The best way to make clear the need for an observational
calibration of theoretical results is given in fig. 2, which
discloses recent theoretical predictions for the MS location of
low mass stars with solar metallicity. It appears that for a given
effective temperature the predicted luminosity can vary over a
range of about $\Delta$logL$\sim$0.12. This means an intrinsic
uncertainty of 0.3 mag, if no additional errors in transferring
luminosity to magnitude are made.

\begin{figure}
\plotfiddle{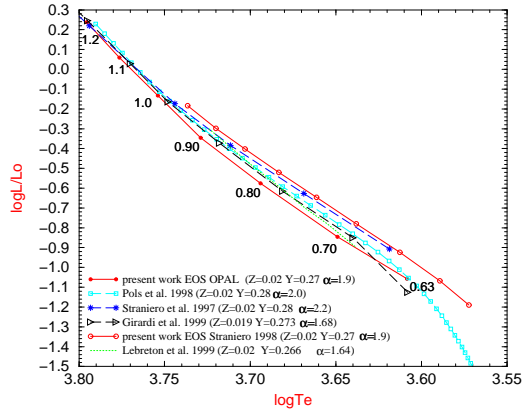}{5.cm}{0}{40}{40}{-110}{-30} \caption{The
predicted CM diagram location for MS stars of solar metallicty as
given by papers appearing in the recent literature.}
\end{figure}

However, even when dealing with models calibrated on the Sun,
there are additional uncertainties. Models in a rather large
portion of the MS are indeed sensitive to the assumptions to be
made about the mixing length parameter which governs the
efficiency of superadiabatic convection in the external stellar
layers (see, e.g., fig.2 in Castellani, Degl'Innocenti \& Prada
Moroni 2001). In principle, the MS location should also depend on
the efficiency of element diffusion: fortunately Salaris,
Groenewegen  \& Weiss (2000) have already found that such a
dependence is largely negligible.

As for the mixing length, one can adopt the value calibrated on
SSM, but there is no reason for the mixing length remaining
constant in varying the star mass, evolutionary phase and chemical
composition! Moreover, the often used calibration of the mixing
length on solar models without element diffusion is not completely
meaningful, since one is using a "wrong Sun" to find out a mixing
length value which pastes the efficiency of convection (actually
larger in SSM) with the effect of sedimentation. Thus at the
present time, this uncertainty cannot be fully removed.

As for the TO luminosity and its dependence on the cluster age,
this is little affected by the mixing length, but the main problem
is wether or not in Pop.II stars element diffusion is at work. It
has indeed already proved that a solar-like diffusion sensitively
affects the predicted evolutionary track of low mass stars. The
lack of evidence for depletion of Fe or Li in the atmosphere of
globular cluster subgiants (Gratton et al 2001, Bonifacio et al
2002) led some people to suspect that, for unknown reasons, Pop.II
star are not affected by this mechanism. Without entering in such
a discussion, one can only notice that we have no observational
evidence concerning the diffusion of He, which - in turn - has the
major influence on the off-MS evolution (Castellani et al 1997).

Passing to the Red Giant phase, one has to notice that these
structures are now sensitive to the efficiency of additional
physical mechanisms, like electron conduction or neutrino
production(for a recent discussion see Salaris, Cassisi \& Weiss
2002) . Whereas color predictions are again depending on the
assumption on the mixing length, the luminosity distribution
appears as a rather firm theoretical result, which has been
already proved in beautiful agreement with observational
constraints. Till recent time, the main uncertainty was related to
electron conduction, whose efficiency governs the mass of the He
core at the flash and, in turn, the predicted luminosity of the
following Horizontal Branch phase.

Old evaluations of electron conduction as given by Hubbard \&
Lampe (1969) have been largely used in stellar models. However
Itoh and coworkers (Itoh et al 1983, 1993, Mitake et al 1984)
presented new and more accurate evaluations but within a range of
validity not properly adequate for RG structures. Owing to this
situation, the choice between the two alternatives has been
largely a matter of opinion of the various authors. Note that
authors adopting Hubbard \& Lampe derives smaller cores and, thus,
lower HB luminosities. More recently, according to updated
evaluations covering the RG internal structures (Pothekin 1999,
Pothekin et al 1999)) one finds that Itoh's choice should have
been preferred, as shown in Table 1 where the He core mass and the
star luminosity at the He flash are reported under the various
labelled assumptions about electron conduction (Piersanti, Prada
Moroni \& Straniero 2002). A detailed discussion on the influence
of other physical ingredients on the HB luminosity can be found in
Cassisi et al (1998), Castellani (1999).

\begin{table}
\caption{Mass of the He core and luminosity at the He flash for a
star of 0.8 M$_\odot$, Z=10$^{-4}$ adopting electron conduction
from the labelled authors }
\begin{center}
\begin{tabular}{ccc}
\tableline
Authors  & Mc & Ltip\\
\tableline
H\&L & 0.5077  & 3.271 \\
Itoh & 0.5116  & 3.292 \\
Pothekin & 0.5108 & 3.289 \\
\tableline \tableline
\end{tabular}
\end{center}
\end{table}

Let me finally quote the still large uncertainties in the HB
lifetimes, often used to calibrate the number ratio N$_{HB}$/
N$_{RG}$ in terms of the original He content of cluster stars. To
follow the time evolution of the central convective cores, one has
to make a choice between classical semiconvection or core
overshooting. In the former case, semiconvection, the uncertainty
on the $^{12}C(\alpha, \gamma)^{16}O$ reaction produces an
uncertainty of about 10\% in the lifetime of central He burning
structures, which increases when the reaction cross section
increases (Cassisi et al 1998, Zoccali et al 2000). HB models with
overshooting taken from the literature can have lifetime even
larger by about 40\%.

\section {Theoretical RR Lyrae}

When dealing with the canonical portion of the CM diagram, let me
spend only few words also on theoretical models for pulsating RR
Lyrae stars, which appears to me a branch of the evolutionary tree
laden with fruit. As well known, evolutionary models give for
these stars a prediction about the mass M and the luminosity  L,
fully constraining the curve of light (period, amplitude.. shape)
predicted by the pulsational theory for each given value of
effective temperature Te. Thus, the coupling of evolutionary with
pulsational theory is a quite efficient   device to submit
theoretical predictions to stringent observational tests.

In this context, it has been already shown that pulsational
constraints can play a relevant role in our knowledge of globular
clusters. As an example, let me only quote  some recent results:
i) the periods of RR Lyrae stars at the Blue Edge for instability
can gives reliable GC distances (Caputo et al. 2000); ii) the
period-K magnitudes relation can be used to give the parallax of
RR Lyr itself better than HST astrometric measurements (Bono et
al. 2002); iii) the shape of the light curve can constrain the
absolute magnitude of the pulsators (Bono et al 2000). Here one
can add that pulsational constraints for RR Lyrae stars in M3
provide pulsator masses in beautiful agreement with evolutionary
predictions (Di Crescenzio, Marconi \& Caputo 2002).

Finally, let me close this brief section with a warning: the mean
V magnitude of a RR Lyrae is not necessarily the magnitude of the
static structure, since it can be fainter up to 0.12 mag for the
fundamental pulsators with large amplitudes (Bono, Caputo \%
Stellingwerf 1995, Di Crescenzio, Marconi \& Caputo 2002). As a
corollarium, it follows that the mean magnitude of a sample of RR
Lyrae stars IS NOT the mean magnitude of the HB.

\section {The Newcomers}

The evidence brought to the light by HST has recently stimulated a
large amount of theoretical investigations on VLM MS stars. The
theoretical approach to these cool and dense structures has
required a remarkable  amount of work to produce adequate
evaluations of  both the EOS and the radiative opacity, as well as
new and adequate  model atmospheres. As expected, stellar models
from different authors but with the same (updated) input physics
give same results, being in beautiful agreement with observation
for the lower metallicity (Cassisi et al 2000 and references
therein). However, for unknown reasons this agreement decreases
increasing the metallicity, becoming worse at solar metallicity
(possibly for problems with the bolometric correction BC ?).

As a warning, it has been recently found  that models at the faint
end of the sequence dramatically depends on the treatment of EOS.
Available EOS tables give the various physical quantities for
selected values of the plasma temperature (T) and pressure (P),
with steps $\Delta$logT=0.08, $\Delta$logP=0.2. However,
decreasing the step (Cassisi, private communication) the models
undergo relevant variations.

\begin{figure}
\plotfiddle{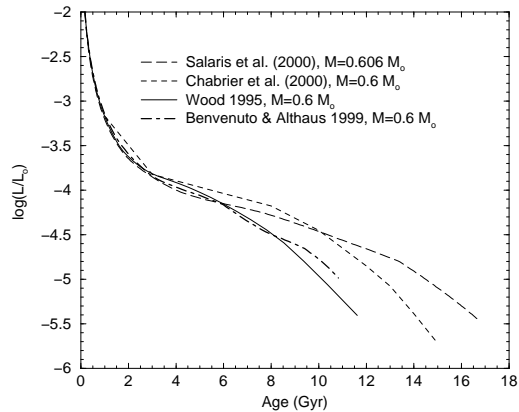}{5.cm}{0}{40}{40}{-110}{-30} \caption{Cooling
laws (luminosity versus time) for CO WD from the current
literature.}
\end{figure}

Finally, as for WD, evaluation of the cooling history obviously
requires a lot of physics, including a detailed treatment of
liquefaction and  crystallization (see, e.g., Prada Moroni \&
Straniero 2002). Looking at the relevant differences among recent
models, one concludes that before using WD as a clock a decision
has to be taken concerning the better physics.

\section{Conclusion}

As a first conclusion, it appears to me that differences among the
recent models of low mass stars are sometime much larger than the
unavoidable uncertainties still connected with the input physics.
A co-operative effort among evolutionary people could greatly
improve this situation, offering to the "common users" of
theoretical predictions a clearer insight on the theoretical
scenarios. In this context, I very much appreciated the
collaboration with the Padua group (L.Girardi) giving light on the
different results concerning   the luminosity of He burning
giants.

In my feeling, users of stellar models should not take theoretical
results without seeing it first.  In the same time, theoretical
people should realize a "code of conduct", including - e.g.- the
prescriptions for observational tests not only with the sun but
also with stellar clusters relevant for the computations. A
procedure often but not always adopted. In this context, stellar
models should be regarded as a bottle of wine: without a label
exhaustively reporting the kind of wine, the vintage an so on, no
one is tempted to taste the contents.

\section {The open question of Tilted HB}

Before closing, I have however to quote a still open question
concerning the basic goal of stellar evolution, as represented by
the lack of explanation for the "tilted" red HB in metal rich GCs.
A possible connection with the dynamic of stars in that dense
clusters has been often suspected, but no clear answer has been
till now reached.

On the contrary, the occurrence of an extended HB in these as in
other GCs is not, strictly speaking, an evolutionary problem, as
it can be easily understood in terms of a sample of RG loosing a
substantial fraction of their mass.

If this is the result of close encounters during the core
collapse, as I suspect, this would imply that post core collapse
clusters develop an EHB lasting  1 - 2 10$^8$ years before coming
back to a normal cluster with a normal HB. However, for the moment
this is only matter of speculation…

\end{document}